\documentclass[prb,twocolumn,showpacs,superscriptaddress,preprintnumbers,amsmath,amssymb]{revtex4}
\usepackage{graphicx}

\usepackage{dcolumn}
\usepackage{bm}

\begin{document}

\newcommand*{\cm}{cm$^{-1}$\,}
\newcommand*{\Th}{T$_{HO}$\,}
\newcommand*{\URS}{URu$_2$Si$_2$\,}


\title{Hybridization gap versus hidden order gap in URu$_2$Si$_2$ as revealed by optical spectroscopy}

\author{W. T. Guo}
\author{Z. G. Chen}

\affiliation{Beijing National Laboratory for Condensed Matter
Physics, Institute of Physics, Chinese Academy of Sciences,
Beijing 100190, China}

\author{T. J. Williams}
\author{J. D. Garrett}
\affiliation{Department of Physics and Astronomy, McMaster
University, Hamilton, Ontario L8S 4M1, Canada}
\author{G. M. Luke}
\affiliation{Department of Physics and Astronomy, McMaster
University, Hamilton, Ontario L8S 4M1, Canada}
\affiliation{Canadian Institute for Advanced Research, Toronto, Ontario, Canada}
\author{N. L. Wang}
\affiliation{Beijing National Laboratory for Condensed Matter
Physics, Institute of Physics, Chinese Academy of Sciences,
Beijing 100190, China}

\begin{abstract}
We present the in-plane optical reflectance measurement on single
crystals of URu$_2$As$_2$. The study revealed a strong
temperature-dependent spectral evolution. Above 50 K, the low
frequency optical conductivity is rather flat without  a
clear Drude-like response, indicating a very short transport life
time of the free carriers. Well below the coherence temperature,
there appears an abrupt spectral weight suppression below 400
cm$^{-1}$, yielding evidence for the formation of a hybridization
energy gap arising from the mixing of the conduction electron and
narrow f-electron bands. A small part of the suppressed spectral
weight was transferred to the low frequency side, leading to a
narrow Drude component, while the majority of the suppressed
spectral weight was transferred to the high frequency side
centered near 4000 cm$^{-1}$. Below the hidden order temperature,
another very prominent energy gap structure was observed, which
leads to the removal of a large part of the Drude component and a
sharp reduction of the carrier scattering rate. The study revealed
that the hybridization gap and the hidden orger gap are distinctly
different: they occur at different energy scales and exhibit
completely different spectral characteristics.

\end{abstract}



\maketitle

\URS has attracted considerable interest in the last two decades
due to its intriguing physical properties and multiple phase
transitions.\cite{Palstra,Schlabitz,Maple,Mydosh} \URS shows
 behavior typical  of heavy fermion metals: its resistivity
increases slightly with decreasing temperature, reaches a maximum
at about 70 K, then decreases fast below 40-50 K.\cite{Palstra}
The behavior is  consistent with the expectation for the
crossover of U 5f electrons from localized behavior with the
formation of local magnetic moments to itinerant transport in
Kondo lattice. In this picture the conduction electrons change from a state of
experiencing strong scattering from local moments of U 5f
electrons to a heavy Fermi liquid state by screening the moment
through a singlet coupling at low temperature. At \Th=17.5 K, the
compound shows a prominent second order phase transition
characterized by a large jump in heat
capacity.\cite{Palstra,Schlabitz,Maple,Mydosh}  Despite many years of study and
countless theoretical proposals, the order
parameter remains unknown. It is therefore referred to as "hidden
order" (HO) state. A tiny magnetic moment formation (0.03
$\mu_B$/U) was determined \cite{Broholm,Mason}, but it is too
small to explain the large entropy loss seen in the specific heat.
 Suggestions for the mysterious order include charge and/or spin density wave formation
 (CDW/SDW)
\cite{Maple,Ikeda,Mineev,Elgazzar} from itinerant U 5f electrons
or multipolar \cite{Santini,Kiss,Haule,Cricchio,Harima} or orbital
orderings \cite{Chandra} in terms of localized 5f electrons, but
no consensus has yet been reached.  Of further interest, a superconducting
transition occurs at much lower temperature T$_c$=1.5 K.

A number of experiments indicate that the HO phase transition is
accompanied by the formation of partial energy gap or
reconstruction of the Fermi surface
(FS).\cite{Maple,Bonn,Degiorgi,Wiebe,Schmidt,Aynajian} However,
some other recent spectroscopic measurements revealed that the
energy gap does not close at the HO transition temperature \Th,
but persist to a higher temperature near 25-30
K.\cite{AynaLiujian,Park,Levallois} A hidden order pseudogap state
has been suggested and examined both experimentally and
theoretically \cite{Haraldsen}. In principle, the HO energy gap
could be obscured by the formation of a hybridization gap
\cite{Dordevic}, a typical feature of heavy fermion metals. This
energy gap arises from the hybridization of conduction electrons
with the flat band of heavy f-electrons in the Kondo lattice Fermi
liquid state. The controversial information about energy gaps
yielded by different experiments could be a result of mixing of
multiple energy gaps. It is essential to have a clear picture
about the energy gaps in order to understand the complex behavior
of \URS. In particular, one should clarify how the hidden order
energy gap is related to the hybridization gap, and whether a HO
pseudogap is actually present?

Optical spectroscopy is a powerful technique to investigate charge
dynamics and band structure of materials as it probes both free
carriers and interband excitations. In particular, it yields
direct information about formation of energy gaps. Infrared
spectroscopy studies on \URS were reported by several groups
previously.\cite{Bonn,Degiorgi,Levallois,Nagel} Early measurements
clearly revealed a partial energy gap formation below the HO
phase transition,\cite{Bonn,Degiorgi} but those studies did
not address possible hybridization energy gaps. On the other hand,
a recent infrared study indicated that there is a pseudogap gap
formation below 30 K in \URS.\cite{Levallois} The feature was
suggested to be a precursor of the HO phase transition at 17.5 K.
In this work, we present a systematic study of single crystals of
URu$_2$Si$_2$. We show that both the hybridization gap and the HO
gap are present in the spectral data, but they occur at different
energy scales and exhibit completely different spectral
characteristics. Our measurement indicates that the peculiar
spectral feature just above the HO temperature is related to the
hybridization energy gap and cannot be taken as a precursor
of HO phase transition. No pseudogap specific to the HO phase
transition is detected.

Single crystals of \URS were grown by the Czochralski method,
using a continuous gettered tri-arc furnace under Ar gas starting
from stoichiometric amounts of the constituent materials. No additional heat treatment
was performed. The
in-plane dc resistivity from 2 to 300 K, measured by a standard
four-probe method in a Quantum Design Physical Properties
measurement system (PPMS) is shown in Figure 1. In agreement with
previous work, the resistivity
increases slightly with decreasing temperature from 300 to 70 K,
then decreases somewhat with further decreasing temperature. A
much faster decrease could be seen below 40$\sim$50 K. The sharp
anomaly at 17.5 K shown in the expanded region in the inset is
attributed to the HO phase transition.

\begin{figure}
\includegraphics[width=7cm,clip]{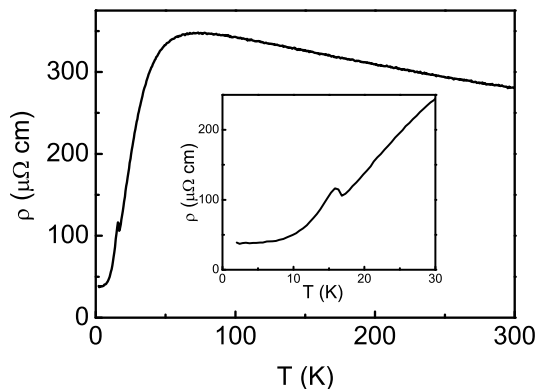}
\caption{The temperature dependent in-plane resistivity of single
crystal \URS. Inset: the expanded plot of $\rho$(T) in the low
temperature range.}
\end{figure}

We performed ptical measurements  on a cleaved in-plane
surfaces of  \URS\ using Bruker 113v, Vertex 80v and a grating-type
spectrometers in the frequency range from 17 to 50000 cm$^{-1}$.
An \textit{in situ} gold and aluminium overcoating technique was
used to get the reflectance \emph{R}($\omega$).\cite{Homes} We
obtained the
real part of conductivity $\sigma_1(\omega)$  by the
Kramers-Kronig transformation of \emph{R}($\omega$). A
Hagen-Rubens relation was used for low frequency extrapolation and a
$\omega^{-1}$ dependence was used for high frequency extrapolation
up to 300000 \cm, above which a $\omega^{-4}$ dependence was
employed.

\begin{figure}
\includegraphics[width=7.3 cm,clip]{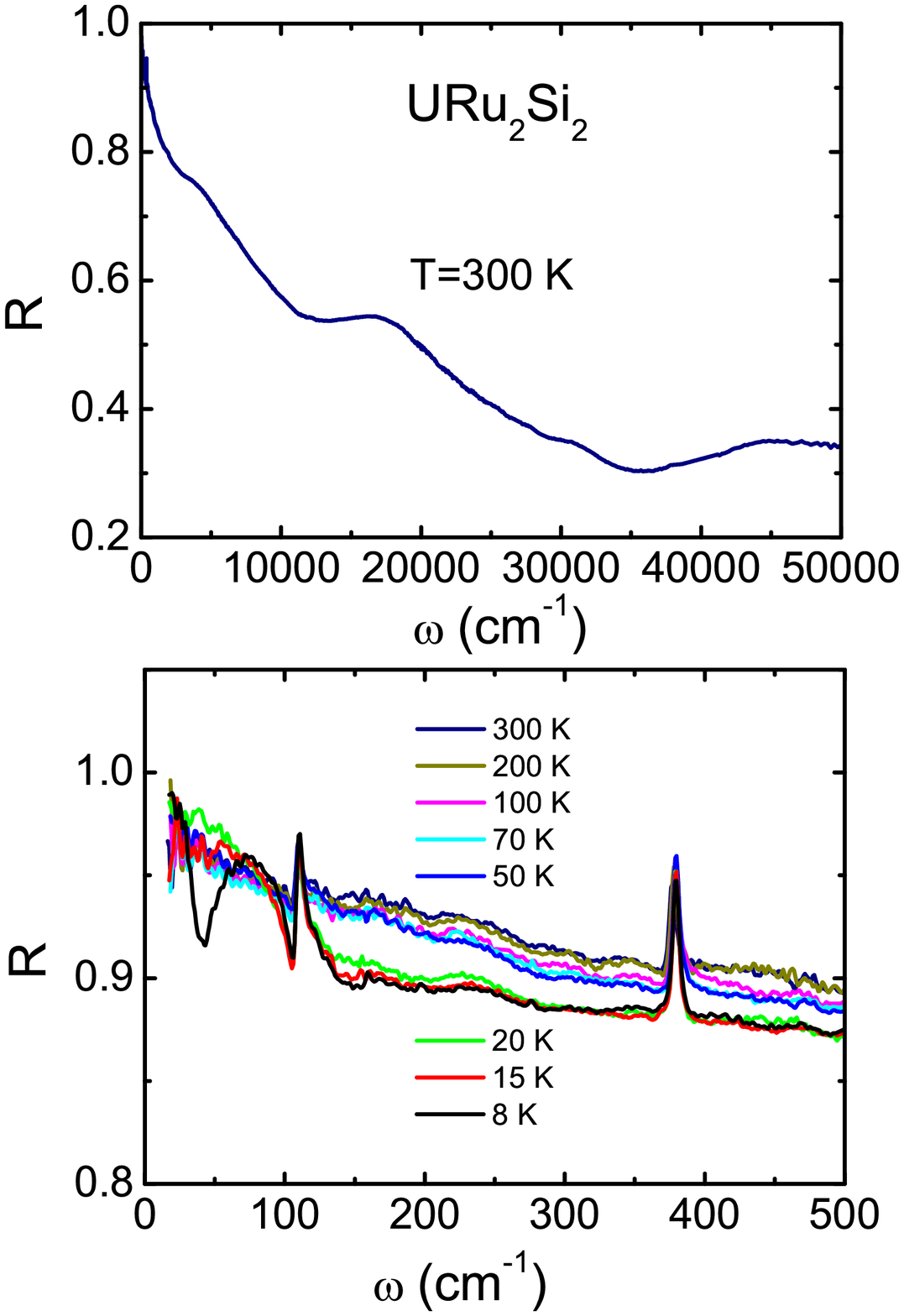}
\caption{\label{fig:EuLaR} Upper panel: the reflectance
R($\omega$) at 300 K for \URS over a broad frequencies up to 50000
\cm. Lower panel: The expanded plot of R($\omega$) in the low
frequency region at different temperatures.}
\end{figure}

Figure 2 shows the reflectance spectra R($\omega$) of \URS for several
different temperatures. The upper panel shows the R($\omega$) at
room temperature up to 50000 \cm, the lower panel shows the
temperature-dependent spectra in an expanded low frequency region
within 500 \cm. Two step-like features could be clearly seen near
3500 \cm ($\sim$0.44 eV) and 16000 \cm ($\sim$2 eV) in R($\omega$)
in the upper panel. They lead to two broad peaks in conductivity
spectrum $\sigma_1(\omega)$ as displayed in the inset of Figure 3,
which we attribute to  interband transitions. Very intriguing
temperature-dependent spectra are seen at low frequencies as shown
in the lower panel of Fig.~2. The reflectance values decrease slightly with
decreasing temperature from 300 K to 50 K, indicating a
non-metallic temperature-dependent response, which is consistent
with the dc resistivity behavior. Much clearer differences in
R($\omega$) are seen from 50 K to 20 K. The reflectance value above
130 \cm is obviously lower than that at higher temperatures.
However,  R($\omega$) shows an upturn  below  about 130 \cm so
that the very low-frequency R($\omega$) values exceed those at
higher temperatures. Below the HO phase transition, another
pronounced spectral feature appears: The R($\omega$)
spectrum is severely suppressed below 60 \cm, but increases
steeply at  ($\omega\leq30$~\cm) lower frequencies, leading to a pronounced dip
feature. This structure is clearly seen in R($\omega$) at
8~K and is still present at
15~K.

The evolution of the electronic states is more clearly
reflected in the conductivity spectra. Figure 3 shows the
$\sigma_1(\omega)$ spectra at different temperatures. The upper
panel shows the $\sigma_1(\omega)$ up to 6000 \cm, the lower panel
shows the spectra in the expanded low frequency region below 500
\cm. The inset shows  $\sigma_1(\omega)$ up to 30000 \cm at
room temperature; the two interband transition peaks can be
clearly seen. There are a number of important temperature-induced
spectral features. First, for temperature higher than 50 K, the
Drude component is virtually invisible. $\sigma_1(\omega)$ show
little temperature dependence below 2000 \cm at room temperature.
A slight decreasing tendency with decreasing frequency is seen at
lower temperature. As is well known, the width of Drude peak is
linked with the scattering rate (or inverse of the transport
lifetime) of the quasiparticle; the measurement result indicates
that there are no well-defined quasiparticles with sufficiently
long transport lifetimes above 50 K, indicating that the conduction electrons
experience extremely strong scatterings from the local U 5f moments.
 Consistent with the dc resistivity
measurement, the low frequency conductivity is suppressed with
decreasing the temperature from 300 K to 50 K. The suppressed
spectral weight below 1500 \cm is transferred to higher-$\omega$
region centered at about 3500 \cm.

\begin{figure}
\includegraphics[width=7.5 cm,clip]{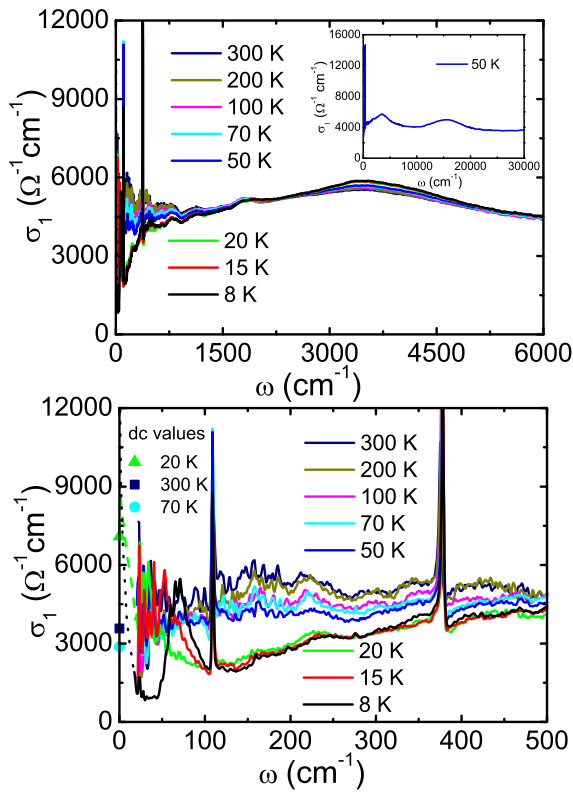}
\caption{\label{fig:EuLaR} Upper panel: the conductivity spectra
$\sigma_1(\omega)$ of \URS at different temperatures. The inset
shows the spectrum at 50 over a broad frequencies up to 30000 \cm.
Lower panel: The expanded plot of $\sigma_1(\omega)$ in the low
frequency region. The dc conductivity values at three temperatures
300, 70 and 20 K were added.}
\end{figure}

Second, in the deeply coherent state, for example at 20 K, there
appears an abrupt spectral weight suppression below 400 \cm in
conductivity spectrum, so that the spectral curve separates itself
clearly from the spectra at temperatures above 50 K. Similar
spectral shapes were observed in two recent
works.\cite{Nagel,Levallois} A small part of the suppressed
spectral weight was transferred to the low frequency side, leading
to the formation of a narrow Drude component, while the majority
of the suppressed spectral weight is still transferred to the
high-$\omega$ side centered near 3500 \cm. The sharp suppression
of $\sigma_1(\omega)$ below 400 \cm is an indication of the
development of an energy gap. The energy level at which the
conductivity reaches the lowest value could be taken as the gap
amplitude, which is about 130 \cm ($\sim$16 meV). This energy gap
is associated with the development of the coherent metallic state
due to the mixing of conduction electron band with the flat 5f
electron band, it is therefore the hybridization energy gap. The
narrow Drude component is derived from the heavy quasiparticles.
The rapid increasing feature is already beyond the low frequency
limit in our measurement and in the extrapolation region.
Nevertheless, the extrapolated curve is verified from a reasonably
good match between the extrapolated zero-frequency limit and the
dc conductivity value. The small spectral weight of the Drude
component is attributed to the heavy quasiparticle effective mass.

The pronounced peak near 3500~\cm should originate predominantly
from the interband transitions shown in the inset of Fig.~3. However, the temperature-induced
spectral weight transfer from low frequency to this energy level
must have a different origin. One possibility is that the spectral
enhancement at such high energy comes from the incoherent part of
the quasi-particle spectral function driven by the on-site Coulomb
repulsion energy (Hubbard U).\cite{Georges,Rozenberg} The presence of
incoherent structures at intermediate frequencies of the order
U/2 to U is supported by dynamical mean-field theory
calculations.\cite{Georges} Nevertheless, the energy scale to which
the spectral weight is transferred depends on the individual compound.
For another prototype HF compound, CeCoIn$_5$, the suppressed
spectral weight is transferred to only around 700 \cm,
\cite{Singley} much lower than the energy scale we infer for  \URS.

The third important feature is the observation of the
prominent energy gap structure below the hidden order transition
temperature \Th\ which  leads to the removal of a large
part of the Drude component. Consistent with previous optical
measurements, a further narrowed Drude component is left in the
extrapolation region. Since the width of the Drude component is the
quasiparticle scattering rate, the opening of the partial energy
gap also leads to a sharp reduction of the carrier scattering
rate. It is worth noting that the removed spectral weight is piled
up just above the energy gap, resulting in a sharp coherent
spectral peak at about 65 \cm ($\sim$8 meV), as
seen in earlier reports.\cite{Bonn,Degiorgi} This spectral feature
is very different from the formation of the hybridization energy
gap which induces a spectral weight transfer to much higher energy
levels. This indicates that the HO energy gap is
completely different from the hybridization energy gap. It is also
worth noting that the formation of hybridization gap is a
crossover phenomenon and not associated with a thermodynamic phase transition such
as in the case of the HO gap.

\begin{figure}
\includegraphics[width=7 cm,clip]{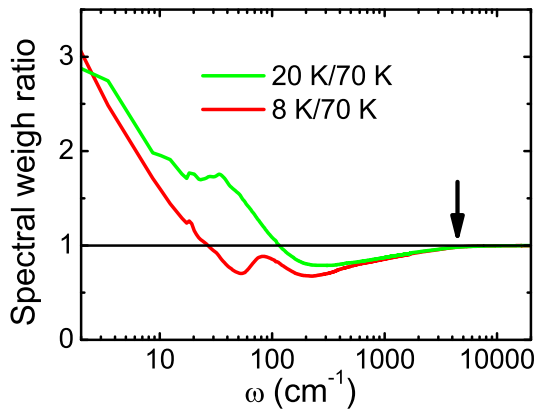}
\caption{(Color online) The normalized spectral weight
SW(T)/SW(70K) up to 10000 \cm. The arrow indicates the energy
scale to which the spectral weight is roughly recovered.}
\end{figure}

The above mentioned spectral weight transfer can be seen more clearly
 in the normalized spectral weight plot shown in Fig. 4. The
spectral weight is defined as
SW=$\int^{\omega_c}_0\sigma_1(\omega)d\omega$, where
 $\omega_c$ is a cut-off frequency. For
clarity we present only the low-temperature normalized spectra
SW(T)/SW(70 K). In the figure, the low-frequency part of SW(20
K)/SW(70 K) is much higher than unity, suggesting the development
of Drude component at low temperature. However, the normalized
spectral weight becomes smaller than unity above 120 \cm due to the
formation of the hybridization gap. It reaches a minimum, then
increases slowly again. The spectral weight is almost fully
recovered near 5000 \cm. Obviously, the spectral weight transfer
related to the hybridization gap formation occurs over a broad
energy scale. In the SW(8 K)/SW(70 K) plot, the spectrum become
less than unity below 30 \cm, reaching a minimum near 50 \cm, then
increasing sharply to almost unity near 80 \cm. This is due to the HO
energy gap formation. The second dip at a higher energy scale and
the gradual recovery up to 5000 \cm could again be attributed to
the spectral weight redistribution associated with the
hybridization energy gap.

The nature of HO remains unknown at present; nevertheless,
the shape of the energy gap provides some hint to this issue. The
pronounced peak in $\sigma_1(\omega)$ is a characteristic
structure of the ``density wave''-type energy gap
excitation.\cite{DresselGruner,Tinkhambook} It arises from the
``case I'' coherent factor effect in the mean-field BCS-like
condensate. In the BCS formalism, the "case I" coherent factor is
for density wave (either charge density wave or spin density wave)
condensate arising from the nesting-induced FS reconstruction, and
"case II" coherent factor is for superconducting
condensate.\cite{DresselGruner,Tinkhambook} For the latter case,
the $\sigma_1(\omega)$ only shows a smooth onset at the energy
gap. For \URS, the shape of the energy gap is almost a "text
book"-like example of "case I" coherent factor, where the location
of the pronounced peak in $\sigma_1(\omega)$ could be identified
as the energy gap, i.e. $2\Delta\approx$ 65 \cm (8 meV). We are
not aware of any other gap-formation mechanism which could produce
such a sharp peak structure in $\sigma_1(\omega)$. We noticed that
recent density-function theory studies have indicated the presence
of two strongly nested FS sheets separated by a nesting vector
$\textbf{Q}_0$ = (0,0,1) in the paramagnetic bct phase.
\cite{Elgazzar,Oppeneer1,Oppeneer2} The nesting wave vector is
identical to the commensurate AF wave vector revealed by the
inelastic neutron scattering experiments. \cite{Broholm,Mason} So,
we believe that the HO is at least related to some type of density
wave associated with the nesting instability of the FS.

The energy scale of the HO energy gap detected here is in good
agreement with the gap energy scale obtained by a number of other
spectroscopic techniques, for example, scanning tunnelling
microscopy (STM) measurement.\cite{Aynajian} The energy scale of
the hybridization energy gap is also in good agreement with that
obtained by recent point contact tunnelling spectroscopy
experiment.\cite{Park} However, the latter measurement failed to
probe the energy gap associated with the HO phase transition at
lower temperature.

Finally, we shall make rough estimates of the mass
enhancement in the heavy fermion coherent state and the change of
the Drude component in the HO state. In principle, the spectral
weight of the Drude component, being equal to the square of the
plasma frequency, could be calculated through the sum rule,
$\omega_p^2$=8$\int^X_0\sigma_1(\omega)d\omega$. The integration
up to X should cover all the spectrum contributed by the free
carriers but still below the inter-band transition. Because the
conductivity spectrum of \URS at high temperature is rather
flat due to very strong scattering from the  U 5f moments,
 it is difficult to separate the Drude component from
the interband transition. Roughly we take X=1500 \cm where we
expect that there is a balance between the Drude component tail
and the onset part of the interband transition, then we get
$\omega_p\approx1.67\times10^4$ \cm for T=300 K. The spectral
weight of the Drude component in the low-temperature coherent
state could be more accurately determined because it separates
distinctly from the remaining part in the conductivity spectra. Taking
X=135 \cm for T=20 K, we get $\omega_p\approx4.28\times10^3$ \cm.
The small plasma frequency in the coherent state is due to the
renormalization of the quasiparticle effective mass arising from
the mixing with the heavy band of U 5f electrons. The mass
enhancement could be obtained from the square of the ratio of the
two plasma frequencies, $m^*/m_B$=[$\omega(300 K)/\omega(20
K)]^2\approx$15. The Drude component is further narrowed due to
the opening of the partial HO gap in the HO state. Taking X=42 \cm
for T=8 K, we get $\omega_p\approx2.09\times10^3$ \cm. The smaller
value relative to that at 20 K is mainly due to the removal of the
FS or the reduction of itinerant carrier density. Assuming the
carrier effective mass does not change below 20 K, the square of
the ratio of the two plasma frequencies, [$\omega(8 K)/\omega(20
K)]^2\approx$0.24, reflects the residual carrier density left in
the HO state. This means that roughly three quarters of FS is
removed by the gapping of the FS. On the other hand, the Drude
component becomes further narrowed. Therefore, the opening of the
partial gap strongly reduces the scattering channel. From all
above analysis, we infer that the HO phase transition is best
understood from the density wave type transition driven by the
Fermi surface instability.

\begin{acknowledgments}
We acknowledge useful discussions with John Mydosh and Yifeng
Yang. This work is supported by the National Science Foundation of
China, and the 973 project of the Ministry of Science and
Technology of China.  Research at McMaster is supported by NSERC and CIFAR.

\end{acknowledgments}

\end{document}